\documentstyle[prl,aps,epsfig]{revtex}
\begin{document}
\draft

\def\overlay#1#2{\setbox0=\hbox{#1}\setbox1=\hbox to \wd0{\hss #2\hss}#1%
\hskip
-2\wd0\copy1}
\twocolumn[
\hsize\textwidth\columnwidth\hsize\csname@twocolumnfalse\endcsname

\title{Slow-light pulses in moving media}
\author{J. Fiur\'{a}\v{s}ek$^{1,2}$, U. Leonhardt$^1$, and R. Parentani$^3$}
\address{$^1$School of Physics and Astronomy, University of St Andrews,
North Haugh, St Andrews, Fife, KY16 9SS, Scotland}
\address{$^2$Department of Optics, Palack\'y University, 17. listopadu 50,
772 07 Olomouc, Czech Republic}
\address{$^3$Laboratoire de Math\'{e}matiques et Physique
Th\'{e}orique, CNRS UPRES A 6083, Universit\'{e} de Tours, 37200
Tours, France}

\maketitle

\begin{abstract}
Slow light in moving media reaches a paradoxical regime when the
flow speed of the medium approaches the group velocity of light.
Pulses can penetrate a seemingly superluminal barrier and are
reflected when the flow slows down.
\end{abstract}
\date{today}

\vskip2pc]

\narrowtext

Imagine a wave packet propagating in a moving medium. Suppose
first that the medium travels as a whole with uniform velocity. In
this case we can construct a co-moving frame of coordinates where
the medium is frozen. The observed speed of the pulse is the
velocity of the wave packet in the co-moving frame transformed
back to the laboratory frame. This mental exercise gives
Einstein's addition theorem of velocities, which, in the case
when both wave and medium propagate much slower than $c$, is
reduced to the Galilean rule that the velocities of wave and
medium simply add up.

Now imagine a non-uniformly moving medium, for example a gas flow
through a nozzle. The addition theorem of velocities is still
valid locally in each infinitesimal piece of the medium.
Consequently, a wave packet moving against the current will never
enter a region where the flow is faster than the wave velocity in
the medium. So, if we send in a pulse of slow light
\cite{Hau,Kash,Budiker,Turukhin,Inouye} to move against the flow
we would expect the pulse to freeze or to run backwards
\cite{Kocharovskaya} when the speed of the medium exceeds the
group velocity $v_g$. Yet exactly the opposite happens. The pulse
will thrive beyond the anticipated superluminal barrier and will
even become faster. On the other hand, the pulse will bounce back
when, paradoxically, the flow slows down.  This effect is
sensitive to minute changes in flow speed and can be applied in
precision measurements of flows with the present state of the art
in slowing down light \cite{Hau,Kash,Budiker,Turukhin,Inouye}.

Let us examine the propagation of slow-light pulses in moving
media. We assume that both the flow speed and the material
parameters of the medium do not change significantly within the
range of an optical wavelength. In this case we can describe light
by the dispersion relation \cite{LPliten} between the wave vector
$\bf{k}^\prime$ and the frequency $\omega^\prime$,
\begin{equation}
k^{\prime 2} -\frac{\omega^{\prime 2}}{c^2}
-\chi(\omega^\prime)\frac{\omega^{\prime 2}}{c^2}=0,
 \label{d1}
\end{equation}
in the locally co-moving frames of the medium indicated with
primes. Slow light is supported by an extremely dispersive medium
where the susceptibility $\chi$ reacts strongly and
proportionally to the detuning between $\omega^\prime$ and a
resonance frequency $\omega_0$ of the medium atoms,
\begin{equation}
\chi(\omega^\prime)=\frac{2c}{v_g}\,\frac{\omega^\prime-\omega_0}{\omega_0}.
\label{chi}
\end{equation}
The group velocity $v_g$ has been demonstrated
\cite{Hau,Kash,Budiker,Turukhin,Inouye} to be as low as a few
meters per second. The specific mechanism to slow down light to
such an impressive degree is not relevant for the analysis
presented here. Slow light can be generated using
Electromagnetically-Induced Transparency \cite{EIT,Harris} or
exploiting light-atom amplification in Bose-Einstein condensates
\cite{Inouye}. Typically \cite{Hau,Kash,Budiker,Turukhin,Inouye},
the steep linear slope (\ref{chi}) of the susceptibility is only
valid within a narrow frequency window
\begin{equation}
|\,\omega^\prime-\omega_0|<\epsilon \frac{v_g}{c}\,\omega_0,
\label{window}
\end{equation}
with $\epsilon$ being in the order of $10^{-3}$. The moving medium
creates a local Doppler shift of the frequency,
\begin{equation}
\omega^\prime=\omega-{\bf u}\cdot {\bf k}, \qquad {\bf
k}^\prime={\bf k}, \label{doppler}
\end{equation}
in the realistic case of non-relativistic flow velocities $\bf u$.

Consider a pulse of slow light with a narrow range of frequencies
$\omega$ near the carrier $\omega_0$. The Doppler detuning
(\ref{doppler}) due to the moving medium is only significant in
the susceptibility. Consequently, we can approximate the
dispersion relation (\ref{d1}) as
\begin{equation}
k^2-k_0^2-\frac{2k_0}{v_g}(\omega-{\bf u}\cdot {\bf
k}-\omega_0)=0.
 \label{d2}
\end{equation}
with
\begin{equation}
k_0=\frac{\omega_0}{c}.
\end{equation}
The group velocity $\bf v$ of the pulse is the derivative of the
frequency $\omega$ with respect to the wave vector $\bf k$, in
analogy to the velocity of a particle with Hamiltonian $\hbar
\omega$ and momentum $\hbar \bf k$. We obtain immediately from the
dispersion relation (\ref{d2})
\begin{equation}
{\bf v}=\frac{\partial \omega}{\partial {\bf k}}
 =  v_g\,\frac{\bf k}{k_0}+ {\bf u}.
 \label{addition}
\end{equation}
This formula is the addition theorem of velocities for slow light
in moving media. Note that the group velocity $v_g$ at resonance
frequency $\omega_0$ and the flow speed $\bf u$ do not simply add
up. The effective group velocity in the co-moving frame is
weighted and directed by the wave vector ${\bf k}$. This has
profound consequences in non-uniformly moving media.

Consider, for simplicity, a one-dimensional model corresponding,
for example, to a gas flow through a tube of varying width.
According to the dispersion relation (\ref{d2}) the wave vector in
1D reduces to
\begin{equation}
k_{\pm}=k_0\left(\pm
\sqrt{1+2\frac{c}{v_g}\delta+\frac{u^2}{v_g^2}}-\frac{u}{v_g}
\right) \label{k}
\end{equation}
with the detuning
\begin{equation}
\delta=\frac{\omega-\omega_0}{\omega_0}.
\end{equation}
The $\pm$ sign indicates propagation with $(+)$ or against $(-)$
the positive current $u$. Consequently, the group velocities are
\begin{equation}
v_{\pm}=\pm v_g\sqrt{1+2\frac{c}{v_g}\delta+\frac{u^2}{v_g^2}}.
\label{v}
\end{equation}
In the case when the medium is moving uniformly with velocity
$u_0$ we can construct a global co-moving frame where the pulse
propagates with group velocity $\pm v_g$ at the carrier frequency
$\omega_0$. When transforming back to the laboratory frame we
must take into account the Doppler shift
\begin{equation}
\delta=\pm \frac{u_0}{c},
 \label{delta}
\end{equation}
and obtain from formula (\ref{v}) the Galilean addition theorem of
velocities,
\begin{equation}
v_0=\pm v_g+u_0.
 \label{v0}
\end{equation}
The situation is completely different when a slow light pulse
propagates in a non-uniformly moving medium. Once launched, the
pulse cannot change its frequency spectrum (unless the flow is not
stationary). Consequently, the pulse cannot adapt to the flow by
continuous Galilei transformations.

Imagine an asymptotically uniform flow with positive speed $u_0$
and a pulse propagating against the current with group velocity
$-v_g+u_0$. When the flow speed changes from the initial $u_0$ to
a varying $u$, the pulse will reach a turning point at a zero of
the group velocity (\ref{v}), {\it i.e.} at
\begin{equation}
u=v_g\sqrt{-2\frac{c}{v_g}\,\delta-1} =
v_g\sqrt{2\frac{u_0}{v_g}-1},
 \label{turn}
\end{equation}
provided that $u_0$ exceeds $v_g/2$ or, equivalently, that the
detuning $\delta$ reaches $-v_g/(2c)$. Note, however, that we have
not yet examined the validity range (\ref{window}) of the
dispersion relation (\ref{d2}). In terms of the effective group
velocity (\ref{v}) we obtain from Eqs.\ (\ref{doppler}) and
(\ref{addition}) the range
\begin{equation}
\left|\, \delta-\frac{u}{v_g}\frac{(v-u)}{c}
\,\right|<\epsilon\,\frac{v_g}{c},
\end{equation}
which gives at the turning point (\ref{turn})
\begin{equation}
|u_0-v_g|<\epsilon v_g.
\end{equation}
Consequently, the initial value of the flow speed should lie in
the vicinity of $v_g$. In this case the velocity (\ref{turn}) at
the turning point is below $u_0$. Pulses propagating against the
current bounce back when the medium reduces speed. This effect
establishes a qualitative change of the behavior of pulses,
triggered by minute variations of the flow velocity, and hence
even the narrow spectral width of slow light
\cite{Hau,Kash,Budiker,Turukhin,Inouye} gives a comfortable
margin for the phenomenon to occur. Figure 1 shows a graphical
method for analyzing slow light in moving media.

\vspace*{-5mm}
\begin{figure}[htbp]
  \begin{center}
    \epsfig{file=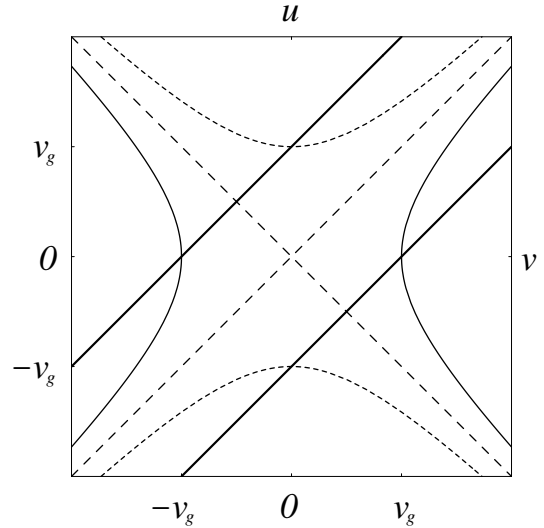,width=0.45\textwidth}
    \caption{Principal regimes of slow-light pulses in moving
    media. The figure shows the group velocity $v$ versus the flow
    speed $u$ for various values of the detuning $\delta$. The
    dotted curves correspond to the common case of zero detuning
    when the pulse cannot freeze to $v=0$. The solid curves
    represent the interesting case, $\delta=-v_g/c$, when the
    pulses reaches turning points outside the group-velocity gap
    between $u=\pm v_g$. The dashed lines show the marginal case
    of $\delta=-v_g/(2c)$. The plots are illustrations of Eq.\
    (\ref{v}) which has, however, a restricted validity range.
    In order to find the working points we
    employ the pair of thick diagonal lines shown in the figure as
    well. At these lines the Doppler detuning $\omega-uk-\omega_0$
    vanishes for one-dimensional wave vectors $k$ satisfying the
    dispersion relation (\ref{d2}), and the addition theorem
    (\ref{addition}) gives $v=\pm v_g+u$. Exactly at the working
    points where the diagonals cross the group-velocity curves the
    Galilean rule of velocity addition is obeyed, but not in a
    close vicinity of the points. This allows a pulse to bounce
    back when the flow slows down, as the figure clearly
    indicates, because the gap between the solid curves lies
    at the side of small medium velocities.}
    \label{fig:figure1}
  \end{center}
\end{figure}

Our effect can serve to sense tiny variations in flow speed, as a
form of slow-light sonar, because the phase of the reflected light
depends on the turning point (\ref{turn}) for a given detuning. In
order to scan a whole range of flow velocities, one should simply
modify the group velocity in Electromagnetically-Induced
Transparency by adjusting the intensity of the coupling beam
\cite{EIT}.

So far, we have analyzed the geometrical optics of slow-light
pulses, the dispersion relation and the group velocity, which
determines the principal effects. Wave optics will tell us the
finer details. In analogy to the correspondence between classical
mechanics and wave mechanics, we regard the frequency $\omega$ and
the wave vector $\bf k$ as differential operators,
\begin{equation}
\omega=i\partial_t, \qquad {\bf k}=-i\nabla,
\end{equation}
and derive from the dispersion relation (\ref{d2}) the wave
equation for the optical field $\varphi$,
\begin{equation}
\left[\left({\bf k}+k_0\frac{\bf u}{v_g}\right)^2
-k_0^2\left(1+2\frac{c}{v_g}\frac{\omega-\omega_0}{\omega_0}
+\frac{u^2}{v_g^2}\right)\right]\varphi=0. \label{wave}
\end{equation}
We see that the flow $\bf u$ acts as an effective vector
potential \cite{LPliten}, similar to the behaviour of waves in
non-dispersive media \cite{Waves}. The flow also generates an
attractive $-u^2/v_g^2$ potential \cite{LPslow} that happens to
be proportional to the Bernoulli pressure of the fluid. When both
medium and light move in only one dimension, say in the $z$
direction, we can safely eliminate the vector potential by the
gauge transformation
\begin{equation}
\varphi=\psi\exp\left(-ik_0 \int \frac{u}{v_g}\, dz\right).
\end{equation}
We arrive at the Schr\"{o}dinger equation
\begin{equation}
\left(\hbar\omega-\frac{\hbar^2 k^2}{2 m}-U\right)\psi=0
\label{sch}
\end{equation}
with the effective mass $m$ and the potential $U$,
\begin{equation}
m=\frac{\hbar \omega_0}{v_g c}, \qquad U=\frac{\hbar\omega_0}{2}
\left(2-\frac{v_g}{c}-\frac{u^2}{v_g c}\right).
\end{equation}
In this way we can benefit from the accumulated experience with
one-dimensional Schr\"{o}dinger waves in scalar potentials. For
instance, we can apply Heller's method \cite{Heller} to analyze
moving wave packets or use the efficient routines available for
numerical simulations. Figure 2 illustrates the two typical
regimes of slow-light wave packets, pulse acceleration and
reflection.

\begin{figure}[htbp]
  \begin{center}
    \epsfig{file=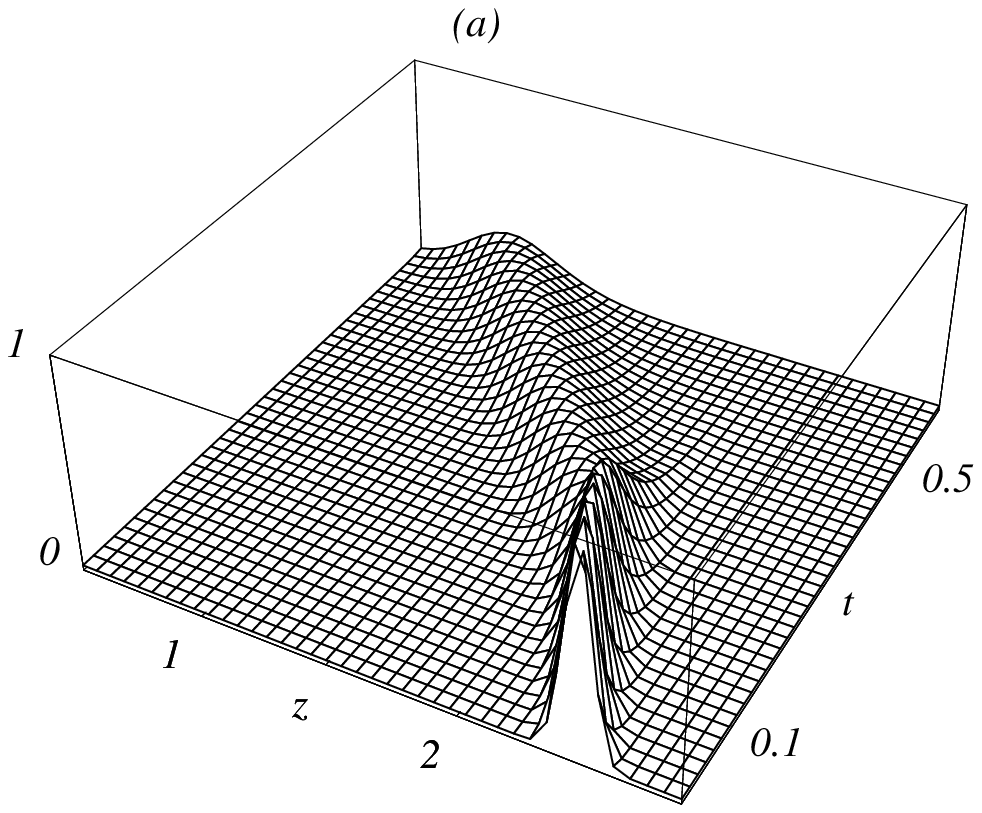,width=0.45\textwidth}
    \epsfig{file=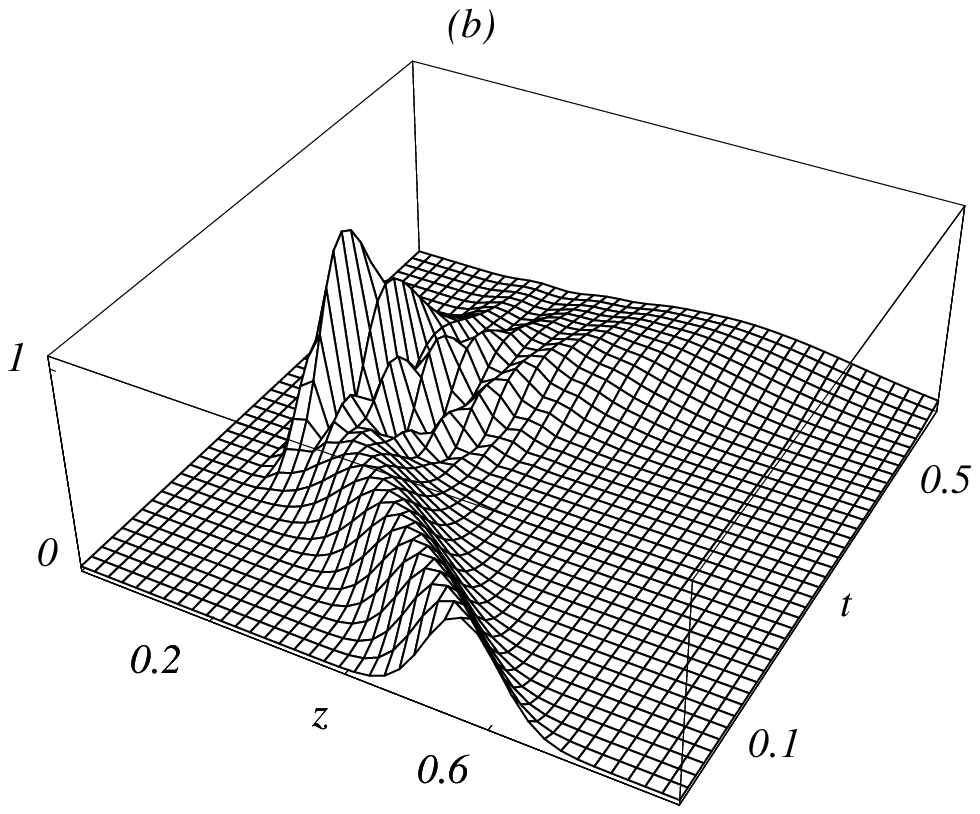,width=0.45\textwidth}
    \vspace*{5mm}
    \caption{
    Propagation of slow-light pulses in a moving medium.
    The medium supports a group velocity of $v_g=300\,{\rm m/s}$ at
    $\omega_0=3\times 10^{15}{\rm Hz}$ and moves with an initial
    flow speed $u_0=0.995 v_g$. The pulses, with resonant
    carrier frequency $\omega=\omega_0(1-u_0/v_g)$ and bandwidth
    $3\,{\rm MHz}$, travel against the flow with a net group
    velocity of $-1.5\,{\rm m/s}$.
    (a) Pulse acceleration. At $z=2\,{\rm mm}$, the flow speed
    increases by $1.15\,{\rm cm/s}$ and the pulse is accelerated to
    reach $-3\,{\rm m/s}$.
    (b) Pulse reflection when the medium velocity decreases.
    At the turning point the flow speed drops by a mere
    $3.8\,{\rm mm/s}$, which illustrates the extreme sensitivity
    of slow-light pulses to variations of the medium velocity.}
    \label{fig:figure2}
  \end{center}
\end{figure}

Figure 3 shows that even a uniform flow can reflect slow-light
pulses when the group velocity $v_g$ is spatially varying, a
regime easily achieved in Electromagnetically-Induced Transparency
\cite{EIT} by modifying the intensity of the coupling beam. At
the turning point the velocity (\ref{v}) of the wave packet
vanishes, which gives, for a uniform flow $u_0$,
\begin{equation}
v_g= -c\delta + \sqrt{c^2\delta^2-u_0^2}. \label{vg}
\end{equation}
For $\delta = -u_0/c$ the pulse is permanently frozen when the
motion of the medium exactly compensates the group velocity. A
slightly different detuning, $\delta < -u_0/c$, will send the
pulse back as soon as $v_g$ reaches the critical value (\ref{vg}).

\begin{figure}[htbp]
  \begin{center}
    \epsfig{file=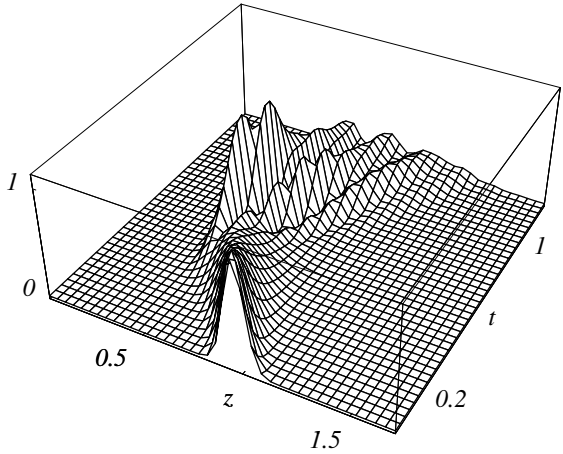,width=0.45\textwidth}
    \vspace*{5mm}
    \caption{Pulse reflection via variations of the group velocity
    $v_g$. The medium supports an initial $v_g$ of $300\,{\rm
    m/s}$ and travels uniformly with $u_0 = 0.995 \times 300\,
    {\rm m/s}$. For a detuning $\delta = -u_0/c$ the pulse is
    frozen when $v_g$ reaches $u_0$. If we set $\delta$ to
    $-1.00001\times u_0/c$ the wave packet bounces back when $v_g$
    drops by $0.16\,{\rm m/s}$, as the figure illustrates.}
  \label{fig:figure3}
  \end{center}
\end{figure}

Finally we remark that similar effects will also occur in a hot
gas with a broad thermal spread of velocities, provided that the
two beams involved in Electromagnetically-Induced Transparency,
the coupling and the probe beam, are co-propagating
\cite{Kocharovskaya}. Given a detuning $\delta_c$ of the coupling
beam, only the atoms which are traveling with the right velocity,
$u_0=-c\delta_c$, reach the narrow EIT resonance. Consequently,
slow light in hot gases is extremely sensitive to minute changes
in the fraction of atoms moving at the critical speed $u_0$.

To summarize, slow light in moving media shows a surprisingly
paradoxical behavior when the flow approaches the group velocity
of light. Pulses can not only penetrate a superluminal barrier but
are also reflected when the flow slows down. Apart from
illustrating puzzles with the addition theorem of velocities in
moving media, our findings may lead to the slow-light analog of a
flow sonar and they add an intriguing aspect to the fascinating
field of slow-light optics
\cite{Hau,Kash,Budiker,Turukhin,Inouye,Kocharovskaya,LPliten,Slowlight}.

We are grateful to John Allen and Malcolm Dunn for fruitful
discussions. J.F. acknowledges the support of the Czech Ministry
of Education, grant LN00A015, and U.L. and R.P. thank ICRA/RIO
and FAPERJ of Brazil.

%%%%%%%%%%%%%%%%%%%%%%%%%%%%%%%%%%%%%%%%%%%%%%%%%%%%%%%%%%%%%%%%%%%%%%%%%%


\begin{thebibliography}{99}

\bibitem{Hau}
L. V. Hau, S. E. Harris, Z. Dutton, and C. H. Behroozi, Nature
{\bf 397}, 594 (1999).

\bibitem{Kash}
M. M. Kash, V. A. Sautenkov, A. S. Zibrov, L. Hollberg, H. Welch,
M. D. Lukin, Y. Rostovsev, E. S. Fry, and M. O. Scully, Phys.
Rev. Lett. {\bf 82}, 5229 (1999).

\bibitem{Budiker} D. Budiker, D. F. Kimball, S. M. Rochester,
and V. V. Yashchuk, Phys. Rev. Lett. {\bf 83}, 1767 (1999).

\bibitem{Turukhin}
A. V. Turukhin, J. A. Musser, V. S. Sudarshanam, M. S. Shahriar,
and P. R. Hemmer, arXiv:quant-ph/0010009.

\bibitem{Inouye}
S. Inouye, R. F. L\"ow, S. Gupta, T. Pfau, A. G\"orlitz, T. L.
Gustavson, D. E. Pritchard, and W. Ketterle, Phys. Rev. Lett.
{\bf 85}, 4225 (2000).

\bibitem{Kocharovskaya}
O. Kocharovskaya, Y. Rostovtsev, and M. O. Scully,
arXiv:quant-ph/0001058.

\bibitem{LPliten} U. Leonhardt and P. Piwnicki, Phys. Rev.
Lett. {\bf 84}, 822 (2000).

\bibitem{EIT}
P. L. Knight, B. Stoicheff, and D. Walls (eds.), Phil. Trans. R.
Soc. Lond. A {\bf 355}, 2215 (1997); S. E. Harris, Phys. Today
{\bf 50}(7), 36 (1997); M. O. Scully and M. Zubairy, {\it Quantum
Optics} (Cambridge University Press, Cambridge, 1997).

\bibitem{Harris}
S. E. Harris, J. E. Field, and A. Kasapi, Phys. Rev. A {\bf 46},
R29 (1992).

\bibitem{Waves}
M. V. Berry, R. G. Chambers, M. D. Large, C. Upstill, and J. C.
Walmsley, Eur. J. Phys. {\bf 1}, 154 (1980); R. J. Cook, H.
Fearn, and P. W. Milonni, Am. J. Phys. {\bf 63}, 705 (1995); P.
Roax, J. de Rosny, M. Tanter, and M. Fink, Phys. Rev. Lett. {\bf
79}, 3170 (1997); U. Leonhardt and P. Piwnicki, Phys. Rev. A {\bf
60}, 4301 (1999); F. Vivanco and F. Melo, Phys. Rev. Lett. {\bf
85}, 2116 (2000).

\bibitem{LPslow}
U. Leonhardt and P. Piwnicki, arXiv:physics/0009093, J. Mod.
Optics (in press).

\bibitem{Heller}
E. J. Heller, J. Chem. Phys. {\bf 62}, 1544 (1975); B. M.
Garraway, J. Phys. B {\bf 33}, 4447 (2000).

\bibitem{Slowlight}
S. E. Harris and L. V. Hau, Phys. Rev. Lett. {\bf 82}, 4611
(1999); M. D. Lukin and A. Imamoglu, {\it ibid.} {\bf 84}, 1419
(2000); M. D. Lukin, S. F. Yelin, and M. Fleischhauer, {\it
ibid.} {\bf 84}, 4232 (2000); M. Fleischhauer and M. D. Lukin,
{\it ibid.} {\bf 84}, 5094 (2000); A. B. Matsko, Y. V.
Rostovtsev, H. Z. Cummins, and M. O. Scully, {\it ibid.} {\bf
84}, 5752 (2000); S. E. Harris, {\it ibid.} {\bf 85}, 4032
(2000); M. Artoni, I. Carusotto, G. C. La Rocca, and F. Bassani
(to be published).

\end{thebibliography}
\end{document}